\documentclass[aps,twocolumn,showpacs,preprintnumbers,amsmath,amssymb,superscriptaddress]{revtex4}

\usepackage{graphicx}
\usepackage{dcolumn}
\usepackage{bm}
\usepackage[koi8-r]{inputenc}
\usepackage{latexsym}
\usepackage{amssymb}

\usepackage[english]{babel}

\begin{document}

\title{Derivation of Hyperbolic Transfer Equations from BGK-Equation}

\author{A. Terentyev}

\affiliation{Institut f\"ur Physik, Technische Universit\"at Chemnitz,
  D-09107 Chemnitz, Germany}

\author{Yu. Skryl}

\affiliation{Institute of Mathematics and Computer Science, University of Latvia, Rai\c na bulv. 29, R\=\i ga, LV-1459, LATVIA}

\date{\today}

\begin{abstract}
We use the integral form of the Boltzmann equation which allows us to take into account the memory effects using the initial condition that selects the solutions going to the local equilibrium Maxwell distribution in the $t \to -\infty$ limit. Implementing the relaxation-time approximation for the collision integral (BGK-equation) we present the derivation of the hyperbolic Navier-Stokes and the hyperbolic heat conduction equations in the first order approximation. It is shown that the relaxation time in the obtained hyperbolic equations is the Maxwellian relaxation time. As special case we obtain the telegraph equation for the heat propagation in static medium and  estimate  the relaxation time for the heat conduction in some materials.

\end{abstract}

\maketitle

\section{Introduction}

\label{sec:1}

The hyperbolic equation for the heat conduction (and also for the diffusion) has been  of interest for a long time. It is known that this equation has so-called the relaxation term which takes into account the propagation of a signal with a finite speed. Many authors have dealt with this problem. Cattaneo \cite{cattaneo48,cattaneo58} and Vernotte \cite{vernotte58} were the first who based their analysis on kinetic theory arguments and added the relaxation term to the Fourier equation. The Cattaneo equation or the hyperbolic heat flux has the form
\begin{equation}
\boldsymbol{q}= - \lambda \nabla T - \tau_g \frac{\partial \boldsymbol{q}}{\partial t}, \label{eq:grad}
\end{equation}   
where $\lambda$ is the thermal conductivity and $\tau_g$ is some relaxation time. In our designations $\tau_g$ is the hyperbolic relaxation time, which is usually obtained from the experiment. For example, for nitrogen $\tau_g=10^{-9}$~s and for aluminium $\tau_g=10^{-11}$~s \cite{lykov67}. From Eq. (\ref{eq:grad}) and the continuity equation  one can write the hyperbolic heat conduction equation.

This point of view has been used in several contexts: e.g. in the analysis of waves in thermoelastic media, fast explosions, and a second sound in solids. Wide ranging bibliography can be found in \cite{ackerman86,roger71,joseph89,donald78,glass87}. An excellent monograph on this problem is in Ref. \cite{jou96}.

The problem of derivation of the hyperbolic equation from microscopic equations caused the special interest for a long time. There are several works on the hyperbolic diffusion. One of the first publications appeared in 1935 \cite{davidov35}. In this paper Davidov obtained the hyperbolic equation only for one-dimensional case and for a constant molecular velocity. The similar results on this problem one can find in other works \cite{fock26,monin59,tolubinsky69,mendez97,fort99}. In the publications of Davies \cite{davies54} and Das \cite{das91} on Brownian motion the hyperbolic diffusion equation was obtained from the Fokker-Plank equation. It means that equations for transfer processes of the hyperbolic form can be derived from the underlying kinetic equations. 

For the heat transfer the Grad's model is of great interest  \cite{grad58}, and its thirteen-moment approximation. This theory solves the Boltzmann equation in non-equilibrium situations, and the hyperbolic heat flux and others hyperbolic-like transfer equations can be obtained in the linear approximation \cite{lykov65}. It is shown that the relaxation time in these equations explicitly depend on the collision parameters. However, exact form of Eq. (\ref{eq:grad}) has not been obtained using Grad's method. Instead of Eq. (\ref{eq:grad}) the hyperbolic heat flux in Grad's expansion contains other terms that can be explained by the expansion of the collision integral.

Nevertheless, there is another way for the theoretical investigation of the non-equilibrium processes by constructing the distribution function $f$ which depends on time in the form of a delay functional \cite{zubarev72,balabanian74,zubarev96}, i.e. taking into account memory effects. There is a reduced description of the non-equilibrium system where  the distribution function $f$ is expressed in terms of the first five moments, which are parameters of the hydrodynamic state (concentration $n$, mass velocity $\boldsymbol{u}$ and temperature $T$). This method is used in the well-known Chapman-Enskog approach \cite{chapman70,ferziger72,kerson63}. The function $f$ is sought as a functional of the latter parameters which satisfy the hydrodynamic equations. These solutions are known as normal solutions \cite{grad58,chapman70,zubarev96}. In contrast to the Chapman-Enskog approach, in the work of Zubarev \cite{zubarev72} the Boltzmann equation is expressed in an integral form, where there is the initial condition that provides the selection of the normal solutions. Consequently it has no other solutions. The analogous method was used for the Liouville equation in several works \cite{zubarev71,zubarev96,zubarev70a,zubarev70b,zubarev70c}. For the derivation of the transfer equations with physically realistic time delays one more condition is necessary such as, for example, the relaxation time approach. If we do not use any additional supposition, it is very difficult to choose the physical relaxation time explicitly, while Eq. (\ref{eq:grad}) is approximation, and as it will be shown in this work, the relaxation time approximation. 

In the present paper we obtain the hyperbolic Navier-Stokes and heat conduction equations from the Boltzmann equation in the relaxation time approximation (BGK-equation \cite{bhatnagar54}) using the idea of the delay functional. This approach allows to obtain the expressions for the relaxational fluxes in integral as well as in differential forms. For this result we use the integral form of the Boltzmann equation with an initial condition, and the relaxation time approximation (RTA) for the collision integral.

In Sec. \ref{sec:be} we give a short description of the basic properties of the Boltzmann equation. In Sec. \ref{sec:if} we consider the integral form for the Boltzmann equation and the relaxation time approximation for the collision integral. The expression for the pressure tensor, heat flux and hyperbolic balance equations are obtained in Sec. \ref{sec:fa}. Then we consider the hyperbolic heat propagation in static medium in Sec. \ref{sec:hhf}. Conclusions are given in Sec. \ref{sec:con}.

For the illustration we consider mono-atomic gas and distribution function in the first approximation with respect to small gradients.

\section{The basic properties of the Boltzmann equation}

\label{sec:be}

Let us consider the Boltzmann equation for the single-particle distribution function of the molecules over velocities $\boldsymbol{v}$ in a point $\boldsymbol{r}$ at an instant time $t$
\begin{equation}
\frac{\partial f}{\partial t} + v_{\alpha} \frac{\partial 
f}{\partial r_{\alpha}} = I(f,f). \label{eq:be1}
\end{equation}
The collision integral is given by
\begin{eqnarray}
&&I(f,f)=\int (f'_1 f' - f_1 f)\ g\ \sigma\ d^2\Omega\ d^3 v_1, 
\nonumber \\
&& f'_1=f(\boldsymbol{r},\boldsymbol{v}'_1,t),\ 
   f'=f(\boldsymbol{r},\boldsymbol{v}',t),\nonumber \\ 
&& f_1=f(\boldsymbol{r},\boldsymbol{v}_1,t),\  
   f=f(\boldsymbol{r},\boldsymbol{v},t),\nonumber \\
&& g = |\boldsymbol{v}-\boldsymbol{v}_1|  \label{eq:be2}
\end{eqnarray}
where $\boldsymbol{v},\ \boldsymbol{v}_1$ is the initial velocities,
$\boldsymbol{v}',\ \boldsymbol{v}'_1$ is the final velocities of molecules after pair collision,  $\sigma$ is the scattering cross section, and $\Omega$ is the solid angle .

The parameters of the hydrodynamic state of the system are expressed through the distribution function $f(\boldsymbol{r},\boldsymbol{v},t)$ as
\begin{subequations}
\label{eq:be3-5} 
\begin{eqnarray}
n(\boldsymbol{r},t)=\int f(\boldsymbol{r},\boldsymbol{v},t) d^3v, 
\label{eq:be3}
\end{eqnarray}
\begin{eqnarray}
u_{\alpha}(\boldsymbol{r},t) = \frac{1}{n(\boldsymbol{r},t)} \int v_{\alpha} f(\boldsymbol{r},\boldsymbol{v},t) d^3v, 
\label{eq:be4}
\end{eqnarray}
\begin{eqnarray}
T(\boldsymbol{r},t)=\frac{2 m }{3 k n(\boldsymbol{r},t)} \int \frac{c^2}{2}f(\boldsymbol{r},\boldsymbol{v},t) d^3v,
\label{eq:be5}
\end{eqnarray}
\end{subequations}
where $n(\boldsymbol{r},t)$ is the particle number density, $\boldsymbol{u}(\boldsymbol{r},t)$ is the mass velocity (or hydrodynamic velocity), $T(\boldsymbol{r},t)$ is the temperature, $\boldsymbol{c}=\boldsymbol{v}-\boldsymbol{u}(\boldsymbol{r},t)$ 
is the thermal velocity, $k$ is the Boltzmann constant, and $m$ is the mass of the particle.

Multiplying Eq. (\ref{eq:be1}) by $1$, $v_{\alpha}$, and $c^2/2$, and integrating with respect to $\boldsymbol{v}$ we obtain the known balance equations
\begin{subequations}
\label{eq:be6-8} 
\begin{eqnarray}
\frac{\partial n}{\partial t} + \frac{\partial
}{\partial r_{\alpha}} (n u_{\alpha}) =0, 
\label{eq:be6}
\end{eqnarray}
\begin{eqnarray}
\frac{\partial u_{\alpha}}{\partial t} + u_{\beta} \frac{\partial 
u_{\alpha}}{\partial r_{\beta}} = - \frac{1}{\rho} \frac{\partial 
p_{\alpha \beta}}{\partial r_{\beta}}, 
\label{eq:be7}
\end{eqnarray}
\begin{eqnarray}
\frac{\partial T}{\partial t} + u_{\alpha} \frac{\partial 
T}{\partial r_{\alpha}} = - \frac{2}{3 k n} \bigg(p_{\alpha \beta} \frac{\partial u_{\alpha}}{\partial r_{\beta}} + \frac{\partial 
q_{\alpha}}{\partial r_{\alpha}}\bigg), 
\label{eq:be8}
\end{eqnarray}
\end{subequations}
where $\rho = m n $ is the mass density. The pressure tensor $p_{\alpha \beta}$ and the heat flux $q_{\alpha}$ are defined by the relations
\begin{subequations}
\label{eq:be9-10} 
\begin{eqnarray}
p_{\alpha \beta}=\int m c_{\alpha} c_{\beta} f(\boldsymbol{r},\boldsymbol{v},t) d^3v, 
\label{eq:be9}
\end{eqnarray}
\begin{eqnarray}
q_{\alpha}= \int c_{\alpha} \frac{mc^2}{2} f(\boldsymbol{r},\boldsymbol{v},t) d^3v. 
\label{eq:be10}
\end{eqnarray}
\end{subequations}

\section{INTEGRAL FORM FOR THE BOLTZMANN EQUATION}

\label{sec:if}

Since the Boltzmann equation is very complicated, simpler kinetic equations have been proposed in the literature. A widely used simplified model is the well known the relaxation time approximation (RTA)
 for the collision integral (e. g. see Refs.\cite{jou96,kerson63,ferziger72,bhatnagar54,smith89}). In this case the resulting Boltzmann equation has the form
\begin{equation}
\frac{\partial f}{\partial t} + v_{\alpha} \frac{\partial
f}{\partial r_{\alpha}} = -\frac{f-f^{(0)}}{\tau_r}, \label{eq:if_0}
\end{equation}
where 
\begin{equation}
f^{(0)}= n (\frac{m}{2 \pi k T})^{\frac{3}{2}} e^{-\frac{m}{2 
kT} c^2},
\label{eq:if2}
\end{equation}
is the local equilibrium Maxwell distribution, $\tau_r$ is the relaxation time which is of the order of the mean free time $\tau_0$.

Let us write the function $f$ in the following form: 
\begin{equation}
f=F+f^{(0)}. \label{eq:if1}
\end{equation}
Substituting (\ref{eq:if1}) in Eq. (\ref{eq:if_0}) we obtain the equation for the function $F$
\begin{equation}
\frac{\partial F}{\partial t} + \frac{1}{\tau_r} F = - \frac{D f^{(0)}}{D t} - v_{\alpha} \frac{\partial F}{\partial r_{\alpha}}.
\label{eq:if8}
\end{equation}
where
\begin{equation}
\frac{D}{D t}= \frac{\partial }{\partial t} + v_{\alpha} 
\frac{\partial}{\partial r_{\alpha}} 
\nonumber
\end{equation}
is the time derivative along the trajectory of a particle.

We note that that the hydrodynamic variables $n(\boldsymbol{r},t)$, $\boldsymbol{u}(\boldsymbol{r},t)$ and $T(\boldsymbol{r},t)$, defined by Eqs. (\ref{eq:be3-5}), coincide with the corresponding variables calculated from the local Maxwellian distribution (\ref{eq:if2}). Thus, the first five moments of the function $F=f-f^{(0)}$ equal zero.

To specify a solution of Eq. (\ref{eq:if8}) the idea of Zubarev and Honkin \cite{zubarev72} is used in the following considerations. One assumes that at some time $t_0$ we have $F(t_0)=0$, i.~e. $f(t_0)=f^{(0)}(t_0)$. Then there considers the limiting process $t-t_0 \to \infty$ supposing that the solution goes to the local Maxwell distribution in the $t \to -\infty$ limit. To satisfy the latter conditions Eq.  (\ref{eq:if8}) is represented in the integral form
\begin{equation}
F(t)= \int_{-\infty}^t e^{\frac{1}{\tau_r} (t'-t)} \bigg (-\frac{D f^{(0)}}{D t} - 
v_{\alpha} \frac{\partial F}{\partial r_{\alpha}}\bigg)_{t'} dt', \label{eq:if9}
\end{equation}
where the subscript $t'$ behind the bracket means that the time argument inside the bracket equal $t'$. Thus, Eq. (\ref{eq:if9}) allows us to take into account the memory effects, and to select normal solutions corresponding to an initial condition
\begin{equation}
\lim_{t \to -\infty} F(t) = 0. \label{eq:if10}
\end{equation}

In an assumption of small gradients of hydrodynamical parameters the integral equation (\ref{eq:if9}) can be solved by expanding $F$ in series in the powers of the Knudsen number $K_n$ ($K_n=l_f/\Delta l$, $l_f$ is the mean free path of particle, and $\Delta l$ is the characteristic space scale for the hydrodynamic processes), or iteratively, assuming
\begin{equation}
F^{(n+1)}(t)= \int_{-\infty}^t e^{\frac{1}{\tau_r}(t'-t)} 
\bigg (-\frac{D f^{(0)}}{D t} - 
v_{\alpha} \frac{\partial F^{(n)}}{\partial r_{\alpha}}\bigg)_{t'} dt'.
\label{eq:if11}
\end{equation}
This iterational procedure can be implemented for the consequent determining of different approximations in constructing of normal solutions. 

In fact, one can obtain the same result if we use formally the Zubarev's method \cite{zubarev72} in the RTA. The difference is only in the exponential term where the term $1/\tau_r$ is substituted by $(1+\varepsilon \tau_r)/\tau_r$. The parameter $\varepsilon$ one can consider as a some parameter providing the convergence of Eq. (\ref{eq:if9}). In our case the role of this parameter  plays the exponential term $1/\tau_r$ with $\varepsilon = 0$.

\section{THE FIRST APPROXIMATION}
\label{sec:fa}

As the first approximation in Eq. (\ref{eq:if11}) we set $F^{(0)}=0$ (i.~e. from Eq. (\ref{eq:if1}) $F^{(0)}=f^{(0)}-f^{(0)}=0$). Thus,
\begin{equation} 
F^{(1)}(t)= -\int_{-\infty}^t e^{\frac{1}{\tau_r}(t'-t)} \bigg(\frac{D
f^{(0)}}{D t}\bigg)_{t'} dt',
\label{eq:fa1}
\end{equation}
where the gradients of $F^{(1)}$ are neglected . On the other hand by differentiating Eq. (\ref{eq:fa1}) over time we obtain
\begin{equation}
\frac{\partial{F^{(1)}}}{\partial{t}} + \frac{1}{\tau_r} F^{(1)} = - \frac{D f^{(0)}}{Dt},
\label{eq:fa15}
\end{equation}
i.~e., the first time derivative $\partial{}/\partial{t}$ of $F^{(1)}$ is taken into account. It can be explained by the fact that the integral equation (\ref{eq:if9}) does not have the derivative $\partial{F}/\partial{t}$, which is not considered in the iterational expansion.

\subsection{The Transfer Lows}

Using Eq. (\ref{eq:if2}) we have
\begin{eqnarray}
\frac{Df^{(0)}}{D t} &=& \frac{f^{(0)}}{R T} \bigg [ RT\bigg (\frac{1}{n} \frac{dn}{dt} + \frac{\partial u_{\alpha}}{\partial r_{\alpha}}\bigg) + 
\nonumber \\
&& + c_{\alpha} \bigg (\frac{d u_{\alpha}}{d t} + \frac{1}{\rho} \frac{\partial p}{\partial r_{\alpha}}\bigg) +
\nonumber \\
&& + \bigg(\frac{c^2}{2 R T} - \frac{3}{2}\bigg) \bigg(\frac{dRT}{dt} + \frac{2}{3} RT \frac{\partial u_{\alpha}}{\partial r_{\alpha}}\bigg) +
\nonumber \\
&& + \bigg(c_\alpha c_\beta - \frac{1}{3} c^2 
\delta_{\alpha 
\beta}\bigg) \frac{\partial u_{\alpha}}{\partial r_{\beta}} + 
\nonumber \\
&&+ c_{\alpha} \bigg(\frac{c^2}{2RT}-\frac{5}{2}\bigg)\frac{\partial 
RT}{\partial 
r_{\alpha}} \bigg]
\label{eq:fa2}
\end{eqnarray}
where
\begin{equation}
\frac{d}{dt} = \frac{\partial }{\partial t} + u_{\alpha}
\frac{\partial}{\partial r_{\alpha}},
\nonumber
\end{equation}
$R=k/m$ is the gas constant.

As mentioned above, the function $F$ does not give contribution to the hydrodynamic parameters. Consequently, for the first five moments of the function $F^{(1)}$ the following conditions have to be fulfilled 
\begin{subequations}
\label{eq:fa3a-3c} 
\begin{eqnarray}
\int F^{(1)} d^3 v = 0,
\label{eq:fa3a}
\end{eqnarray}
\begin{eqnarray}
\int v_{\alpha} F^{(1)} d^3 v = 0,
\label{eq:fa3b}
\end{eqnarray}
\begin{eqnarray}
\int \frac{c^2}{2} F^{(1)} d^3 v = 0.
\label{eq:fa3c}
\end{eqnarray}
\end{subequations}
From these conditions there follows that the first three terms of Eq. (\ref{eq:fa2}) in round brackets, which are the hydrodynamical equations of ideal fluid, have to be equal zero. To prove this we will follow the work of Zubarev and Honkin \cite{zubarev72} representing their results for our case. First, we will show that the integrals
\begin{equation}
I_{\alpha \beta}(t) = \int d^3v \psi_{\alpha} \int_{-\infty}^t dt' e^{\frac{1}{\tau_r} (t'-t)} (f^{(0)} \psi_{\beta})_{t'},
\label{a1}
\end{equation}
\begin{displaymath}
\psi_{\beta} = \left \{
\begin{array}{ll}
1, & \beta=0 \\
c_{\beta}, & \beta = 1,2,3 \\
\frac{c^2}{2} - \frac{3}{2} \frac{k}{m} T, & \beta=4
\end{array} \right.
\end{displaymath}
equal zero for $\alpha \ne \beta$, and do not equal zero for $\alpha = \beta$.

Let us consider a function
\begin{equation}
g_{\beta}(t) = \int_{-\infty}^t e^{\frac{1}{\tau_r} (t'-t)} (f^{(0)} \psi_{\beta})_{t'} dt'.
\label{a2}
\end{equation}
The function $g_{\beta}(t)$ satisfies an equation
\begin{equation}
\bigg(\frac{\partial}{\partial t} + \frac{1}{\tau_r} \bigg)g_{\beta}(t) = f^{(0)} \psi_{\beta}.
\label{a3}
\end{equation}
We multiply Eq. (\ref{a3}) on $\psi_{\alpha}$ and integrate it over $\boldsymbol{v}$ obtaining
\begin{equation}
\bigg(\frac{\partial}{\partial t} + \frac{1}{\tau_r} \bigg)I_{\alpha \beta}(t) = J_{\alpha \beta},
\label{a4}
\end{equation}
where
\begin{equation}
J_{\alpha \beta} = \int d^3 v f^{(0)} \psi_{\alpha} \psi_{\beta}.
\label{a5}
\end{equation}
Taking into account that the integrals $J_{\alpha \beta}=0$ for $\alpha \ne \beta$, and $J_{\alpha \beta} \ne 0$ for $\alpha = \beta$, we prove the our statement for the integrals (\ref{a1}).

In similar way one can show that
\begin{subequations}
\label{a7a-7c} 
\begin{eqnarray}
\int d^3v \psi_{\gamma} \int_{-\infty}^t dt' e^{\frac{1}{\tau_r} (t'-t)} (f^{(0)} \varphi_{\alpha \beta})_{t'}=0,
\label{a7a}
\end{eqnarray}
\begin{eqnarray}
\int d^3v \psi_{\gamma} \int_{-\infty}^t dt' e^{\frac{1}{\tau_r} (t'-t)} (f^{(0)} \varphi_{\alpha})_{t'}=0,
\label{a7b}
\end{eqnarray}
\begin{eqnarray}
\int d^3v \varphi_{\alpha \beta} \int_{-\infty}^t dt' e^{\frac{1}{\tau_r} (t'-t)} (f^{(0)} \varphi_{\alpha})_{t'}=0,
\label{a7c}
\end{eqnarray}
\end{subequations}
where
\begin{subequations}
\label{a8a-8b} 
\begin{eqnarray}
\varphi_{\alpha \beta} = c_\alpha c_\beta - \frac{1}{3} c^2 
\delta_{\alpha \beta},
\label{a8a}
\end{eqnarray}
\begin{eqnarray}
\varphi_{\alpha} = c_{\alpha} \bigg(\frac{c^2}{2RT}-\frac{5}{2}\bigg).
\label{a8b}
\end{eqnarray}
\end{subequations}
The relations (\ref{a7a-7c}) follow from the properties of the mutual  orthogonality of the functions $\psi_{\beta}$, $\varphi_{\alpha \beta}$ and $\varphi_{\alpha}$. 

Let us substitute expressions (\ref{eq:fa1}) and (\ref{eq:fa2}) into Eqs. (\ref{eq:fa3a-3c}). Then using the properties of the integrals (\ref{a1}) and (\ref{a7a-7c}) we obtain that the relations (\ref{eq:fa3a-3c}) are fulfilled only in the case when the hydrodynamical equations of ideal fluid are implemented
\begin{subequations}
\label{eq:fa4-6} 
\begin{eqnarray}
\frac{\partial n}{\partial t} + n \frac{\partial 
u_{\alpha}}{\partial r_{\alpha}} =0, \label{eq:fa4}
\end{eqnarray}
\begin{eqnarray}
\frac{\partial u_{\alpha}}{\partial t} + \frac{1}{\rho} \frac{\partial 
p}{\partial r_{\alpha}} = 0, \label{eq:fa5}
\end{eqnarray}
\begin{eqnarray}
\frac{\partial T}{\partial t} + \frac{2}{3} T \frac{\partial u_{\alpha}}{\partial r_{\alpha}} = 0, \label{eq:fa6}
\end{eqnarray}
\end{subequations}
and the function $F^{(1)}$ has the form 
\begin{eqnarray}
F^{(1)} &=&\ -\int_{-\infty}^t e^{\frac{1}{\tau_r}(t'-t)} \times 
\nonumber \\
&\times& \bigg[\frac{f^{(0)}}{RT} \bigg( \varphi_{\alpha \beta} \frac{\partial u_{\alpha}}{\partial r_{\beta}} + 
\varphi_{\alpha} \frac{\partial 
RT(t')}{\partial r_{\alpha}} \bigg) \bigg]_{t'} dt'. \label{eq:fa7}
\end{eqnarray}
We see that the right hand side of Eq. (\ref{eq:fa7}) is expressed in terms of gradients of the hydrodynamic variables. Thus, the function $F^{(1)}$ represents a normal solution of the Boltzmann equation.

We can obtain the pressure tensor and the heat flux in the first approximation using expressions (\ref{eq:be9-10}) and orthogonal properties for $\varphi_{\alpha \beta}$ and $\varphi_{\alpha}$:
\begin{subequations}
\label{eq:fa8-9} 
\begin{eqnarray}
&& p_{\alpha \beta} - p \delta_{\alpha \beta} = -\int_{-\infty}^t e^{\frac{1}{\tau_r} (t'-t)} M(t,t') \times \nonumber \\
&& \times \bigg(\frac{\partial u_{\alpha}}{\partial r_{\beta}} + \frac{\partial u_{\beta}}{\partial r_{\alpha}} - \frac{2}{3} \frac{\partial u_{\gamma}}{\partial r_{\gamma}} \delta_{\alpha \beta}\bigg)_{t'} dt', \label{eq:fa8}
\end{eqnarray}
\begin{eqnarray}
q_{\alpha} = - \int_{-\infty}^t e^{\frac{1}{\tau_r} (t'-t)} 
L(t,t') \frac{\partial T(t')}{\partial r_{\alpha}} dt',
\label{eq:fa9}
\end{eqnarray}
\end{subequations}
where
\begin{subequations}
\label{eq:fa10-11} 
\begin{eqnarray}
M(t,t')=\frac{1}{10} \int m c_\alpha c_\beta \bigg( \frac{f^{(0)}}{RT}  \varphi_{\alpha \beta} \bigg)_{t'} d^3 c, \label{eq:fa10}
\end{eqnarray}
\begin{eqnarray}
L(t,t')=\frac{1}{3} \int c_{\alpha} \frac{m c^2}{2} \bigg( \frac{f^{(0)}}{T} \varphi_{\alpha} \bigg)_{t'} d^3c \label{eq:fa11}
\end{eqnarray}
\end{subequations}
are the kernels of the equations (\ref{eq:fa8-9}). 

The integral form for the heat flux, like Eq. (\ref{eq:fa9}), is known relatively long time \cite{gurtin68,nunziato71}. These equations describe the transfer in the local non-equilibrium systems or in the so-called materials with memory \cite{sobolev91}. However, as shown above the same result one can also obtain for the pressure tensor Eq. (\ref{eq:fa8}).
 
Eqs. (\ref{eq:fa8-9}) take the classical form for $t \gg \tau_r$. In this case the exponent becomes the delta-function, consequently one can carry out behind the integral the other integrands in the point $t'=t$
\begin{subequations}
\label{eq:fa8a-9a} 
\begin{eqnarray}
&& (p_{\alpha \beta} - p \delta_{\alpha \beta}) \bigg|_{t \gg \tau_r}= -M(t) \times 
\nonumber \\
&& \times \bigg(\frac{\partial u_{\alpha}}{\partial r_{\beta}} + \frac{\partial u_{\beta}}{\partial r_{\alpha}} - \frac{2}{3} \frac{\partial u_{\gamma}}{\partial r_{\gamma}} \delta_{\alpha \beta}\bigg) \int_{-\infty}^t e^{\frac{1}{\tau_r} (t'-t)} dt' =
\nonumber \\
&& =  - \tau_r M(t) \bigg(\frac{\partial u_{\alpha}}{\partial r_{\beta}} + \frac{\partial u_{\beta}}{\partial r_{\alpha}} - \frac{2}{3} \frac{\partial u_{\gamma}}{\partial r_{\gamma}} \delta_{\alpha \beta}\bigg) =
\nonumber \\
&& =  - \mu \bigg(\frac{\partial u_{\alpha}}{\partial r_{\beta}} + \frac{\partial u_{\beta}}{\partial r_{\alpha}} - \frac{2}{3} \frac{\partial u_{\gamma}}{\partial r_{\gamma}} \delta_{\alpha \beta}\bigg), \label{eq:fa8a}
\end{eqnarray}
\begin{eqnarray}
q_{\alpha} \bigg|_{t \gg \tau_r} &=& - L(t) \frac{\partial T(t)}{\partial r_{\alpha}} \int_{-\infty}^t e^{\frac{1}{\tau_r} (t'-t)}  dt' =
\nonumber \\
&& = - \tau_r L(t) \frac{\partial T(t)}{\partial r_{\alpha}} = -\lambda \frac{\partial T(t)}{\partial r_{\alpha}},
\label{eq:fa9a}
\end{eqnarray}
\end{subequations}
where $\mu$ and $\lambda$ are the classical shear viscosity and thermal conductivity. Using the Appendix \ref{sec_app2} one can show that they have forms:
\begin{subequations}
\label{eq:bal5a-6a}
\begin{eqnarray}
\mu=\tau_r M(t)= \tau_r n k T = \frac{1}{2} \tau_r \bar{v}^2 n m,
\label{bal5a}
\end{eqnarray}
\begin{eqnarray}
\lambda=\tau_r L(t)= \tau_r \frac{5}{2} \frac{k T}{m} n k = \frac{5}{4} \tau_r \bar{v}^2 n k,
\label{bal6a}
\end{eqnarray}
\end{subequations}
where $\bar{v}=\sqrt{2kT/m}$ is the average molecular velocity.
These equations have the known classical limit:
\begin{equation}
\tau_r \bar{v}^2 \rightarrow const,\ \textrm{if}\ \tau_r \rightarrow 0 \ \textrm{and} \ \bar{v}^2 \rightarrow \infty.
\label{eq:limit}
\end{equation}

It is important to note that in contrast to the known Navier-Stokes and Fourier lows the equations (\ref{eq:fa8-9}) contain time delay.

\subsection{Balance Equations}
\label{sec:bal}

In order to obtain the hydrodynamic balance equations in the first approximation we differentiate Eqs. (\ref{eq:fa8-9}) over time:
\begin{subequations}
\label{eq:bal1-2}
\begin{eqnarray}
\frac{\partial{(p_{\alpha \beta}-p \delta_{\alpha \beta})}}{\partial{t}} = && - M(t) \bigg(\frac{\partial u_{\alpha}}{\partial r_{\beta}} + \frac{\partial u_{\beta}}{\partial r_{\alpha}} - \frac{2}{3} \frac{\partial u_{\gamma}}{\partial r_{\gamma}} \delta_{\alpha \beta}\bigg) -
\nonumber \\
&& - \frac{1}{\tau_r} (p_{\alpha \beta}-p \delta_{\alpha \beta})+\tilde{p}_{\alpha \beta},
\label{eq:bal1}
\end{eqnarray}
\begin{eqnarray}
\frac{\partial q_{\alpha}}{\partial t} = -L(t)
\frac{\partial T(t)}{\partial r_{\alpha}} - \frac{1}{\tau_r} q_{\alpha}+\tilde{q}_{\alpha}. \label{eq:bal2}
\end{eqnarray}
\end{subequations}
where
\begin{subequations}
\label{eq:bal1a-2a} 
\begin{eqnarray}
\tilde{p}_{\alpha \beta} &=& -\int_{-\infty}^t e^{\frac{1}{\tau_r} (t'-t)} \frac{\partial M(t,t')}{\partial t} \times 
\nonumber \\
&& \times \bigg(\frac{\partial u_{\alpha}}{\partial r_{\beta}} + \frac{\partial u_{\beta}}{\partial r_{\alpha}} - \frac{2}{3} \frac{\partial u_{\gamma}}{\partial r_{\gamma}} \delta_{\alpha \beta}\bigg)_{t'} dt', \label{eq:bal1a}
\end{eqnarray}
\begin{eqnarray}
\tilde{q}_{\alpha} = - \int_{-\infty}^t e^{\frac{1}{\tau_r} (t'-t)} 
\frac{\partial L(t,t')}{\partial t} \frac{\partial T(t')}{\partial r_{\alpha}} dt'
\label{eq:bal2a}
\end{eqnarray}
\end{subequations}
are the last terms in Eqs. (\ref{eq:bal1-2}),
\begin{subequations}
\label{eq:bal1b-2b}
\begin{eqnarray}
\frac{\partial M(t,t')}{\partial t} &=& 
-\frac{m}{10} \int \bigg(\frac{\partial u_{\alpha}}{\partial t} c_{\beta} + \frac{\partial u_{\beta}}{\partial t} c_{\alpha} \bigg) \times
\nonumber \\
&& \times\bigg( \frac{f^{(0)}}{RT}  \varphi_{\alpha \beta} \bigg)_{t'} d^3 c, 
\label{eq:bal1b}
\end{eqnarray}
\begin{eqnarray}
\frac{\partial L(t,t')}{\partial t}&=&-\frac{m}{3} \int \bigg(\frac{\partial u_{\alpha}}{\partial t} \frac{c^2}{2} + \frac{\partial u_{\beta}}{\partial t} c_{\alpha} c_{\beta}\bigg) \times 
\nonumber \\
&& \times \bigg(\frac{f^{(0)}}{T} \varphi_{\alpha} \bigg)_{t'} d^3c \label{eq:bal2b}
\end{eqnarray}
\end{subequations}
are the kernels of Eqs. (\ref{eq:bal1a-2a}). Since the expressions  (\ref{eq:bal1b-2b}) contain the orthogonal functions in the meaning of the scalar product (\ref{a1}), the integrals (\ref{eq:bal1a-2a}) equal zero
\begin{equation}
\tilde{p}_{\alpha \beta}=0,\ \tilde{q}_{\alpha} = 0.
\label{eq:bal1c-2c}
\end{equation}

For further considerations we write Eqs. (\ref{eq:bal1-2}) in a more convenient form:
\begin{subequations}
\label{eq:bal3-4}
\begin{eqnarray}
p_{\alpha \beta} = && p \delta_{\alpha \beta} - 2 \mu \bigg(\Lambda_{\alpha \beta} - \frac{1}{3} \frac{\partial u_{\gamma}}{\partial r_{\gamma}} \delta_{\alpha \beta}\bigg) -
\nonumber \\
&& \tau_r \frac{\partial{(p_{\alpha \beta}-p \delta_{\alpha \beta})}}{\partial{t}},
\label{eq:bal3}
\end{eqnarray}
\begin{eqnarray}
q_{\alpha} = - \lambda \frac{\partial T}{\partial r_{\alpha}} - \tau_r \frac{\partial q_{\alpha}}{\partial t},
\label{eq:bal4}
\end{eqnarray}
\end{subequations}
where
\begin{eqnarray}
\Lambda_{\alpha \beta}= \frac{1}{2} \bigg(\frac{\partial u_{\alpha}}{\partial r_{\beta}} + \frac{\partial u_{\beta}}{\partial r_{\alpha}}\bigg).
\label{bal7}
\end{eqnarray}

To obtain the hydrodynamic balance equations, we substitute $p_{\alpha \beta}$ and $q_{\alpha}$ into equations (\ref{eq:be7}) and  (\ref{eq:be8}), which are the conservations laws. First we calculate some of the essential expressions that will be needed later:
\begin{eqnarray}
\frac{\partial{p_{\alpha \beta}}}{\partial{r_{\beta}}}&=&\frac{\partial{p}}{\partial{r_{\beta}}}-\mu \bigg(\nabla^2 u_{\alpha}+\frac{1}{3} \frac{\partial}{\partial{r_{\beta}}} (\boldsymbol{\nabla}\! \cdot\! \boldsymbol{u})\bigg) - 
\nonumber \\
&& - 2 \frac{\partial{\mu}}{{r_{\beta}}} \bigg(\Lambda_{\alpha \beta} -\frac{1}{3} \frac{\partial u_{\gamma}}{\partial r_{\gamma}} \delta_{\alpha \beta}\bigg)-
\nonumber \\
&& - \tau_r \frac{\partial}{\partial t} \bigg(\frac{\partial{p_{\alpha \beta}}}{\partial{r_{\beta}}}-\frac{\partial{p}}{\partial{r_{\beta}}}\bigg),
\label{eq:bal8}
\end{eqnarray}
\begin{eqnarray}
p_{\alpha \beta} \frac{\partial{u_{\alpha}}}{\partial{r_{\beta}}}&=&\bigg(\tau_r \frac{\partial p}{\partial t}+p\bigg) (\boldsymbol{\nabla}\! \cdot\! \boldsymbol{u}) - 2 \mu \Lambda_{\alpha \beta} \frac{\partial{u_{\alpha}}}{\partial{r_{\beta}}}+
\nonumber \\
&&+\frac{2}{3} \mu (\boldsymbol{\nabla}\! \cdot\! \boldsymbol{u})^2 - \tau_r \frac{\partial p_{\alpha \beta}}{\partial t} \frac{\partial{u_{\alpha}}}{\partial{r_{\beta}}}, 
\label{eq:bal9}
\end{eqnarray}
\begin{equation}
\frac{\partial q_{\alpha}}{\partial{r_{\alpha}}}=-\lambda \nabla^2 T - \nabla \lambda \nabla T - \tau_r \frac{\partial}{\partial t} \frac{\partial q_{\alpha}}{\partial{r_{\alpha}}}.
\label{eq:bal_10}
\end{equation}

Taking into account (\ref{eq:be7}) and (\ref{eq:bal8}) we obtain:
\begin{eqnarray}
&& \tau_r \frac{\partial}{\partial t} 
\bigg[\rho \bigg(\frac{\partial u_{\alpha}}{\partial t} + u_{\beta} \frac{\partial u_{\alpha}}{\partial r_{\beta}}\bigg)\bigg] + \rho \bigg(\frac{\partial u_{\alpha}}{\partial t} + u_{\beta} \frac{\partial u_{\alpha}}{\partial r_{\beta}}\bigg) = 
\nonumber \\
&& = - \frac{\partial}{\partial r_{\alpha}}\bigg(p-\frac{\mu}{3}(\boldsymbol{\nabla}\! \cdot\! \boldsymbol{u})\bigg) + \mu \nabla^2 u_{\alpha} - \tau_r \frac{\partial}{\partial r_{\alpha}} \frac{\partial p}{\partial t} + 
\nonumber \\
&& + Z,
\label{eq:bal_11}
\end{eqnarray}
where
\begin{equation}
Z = 2 \frac{\partial{\mu}}{\partial {r_{\beta}}} \bigg(\Lambda_{\alpha \beta} -\frac{1}{3} \frac{\partial u_{\gamma}}{\partial r_{\gamma}} \delta_{\alpha \beta}\bigg).
\nonumber
\end{equation}
From Eqs. (\ref{eq:be8}), (\ref{eq:bal9}) and (\ref{eq:bal_10}) one has:
\begin{eqnarray}
&& \frac{3}{2} k \tau_r \frac{\partial}{\partial t} \bigg[n \bigg(\frac{\partial T}{\partial t} + u_{\beta} \frac{\partial T}{\partial r_{\beta}}\bigg)\bigg] + 
n \bigg (\frac{\partial T}{\partial t} + u_{\beta} \frac{\partial T}{\partial 
r_{\beta}}\bigg) = 
\nonumber \\
&& = \lambda \nabla^2 T + \nabla \lambda \nabla T - 
\nonumber \\
&& - \bigg(\tau_r \frac{\partial p}{\partial t} + p\bigg) (\boldsymbol{\nabla}\! \cdot\! \boldsymbol{u}) + 2 \mu \Lambda_{\alpha \beta} \frac{\partial 
u_{\alpha}}{\partial r_{\beta}} - \frac{2}{3} \mu (\boldsymbol{\nabla}\! \cdot\! \boldsymbol{u})^2 +
\nonumber \\
&& + \tau_r \frac{\partial p_{\alpha \beta}}{\partial t} \frac{\partial
u_{\alpha}}{\partial r_{\beta}} - 
\nonumber \\
&& - \tau_r \frac{\partial}{\partial t} \bigg [\bigg (\tau_r 
\frac{\partial p}{\partial t} +
 p\bigg) (\boldsymbol{\nabla}\! \cdot\! \boldsymbol{u}) - 2 \mu \Lambda_{\alpha \beta} \frac{\partial
u_{\alpha}}{\partial r_{\beta}} + \frac{2}{3} \mu (\boldsymbol{\nabla}\! \cdot\! \boldsymbol{u})^2 -
\nonumber \\
&& - \tau_r \frac{\partial p_{\alpha \beta}}{\partial t} \frac{\partial
u_{\alpha}}{\partial r_{\beta}}\bigg].  
\label{eq:bal_12}
\end{eqnarray}
In Eqs. (\ref{eq:bal_11}) and (\ref{eq:bal_12}) terms of the first order are 
$\mu$, $\lambda$, $u$, $\partial u/\partial x_i$, 
$\partial \rho/\partial x_i$ and $\partial T/\partial x_i$. We keep only the terms of the first order and neglect all terms which contain the derivatives over $\mu$ and $\lambda$. Then in Eq. (\ref{eq:bal_11}) one can neglect $Z$, and, in the right-hand part of Eq. (\ref{eq:bal_12}), the second, fourth, fifth term, and, in the square brackets, the second and third term. For the sixth term, and the last term in the square brackets on the right-hand side of Eq. (\ref{eq:bal_12}) one can apply the following approximation:
\begin{eqnarray}
&& \tau_r \frac{\partial p_{\alpha \beta}}{\partial t} \frac{\partial
u_{\alpha}}{\partial r_{\beta}} = \bigg [\frac{\partial p}{\partial t} 
\delta_{\alpha \beta} - 2 \frac{\partial \mu}{\partial t} \bigg (\Lambda_{\alpha \beta} - \frac{1}{3}(\boldsymbol{\nabla}\! \cdot\! \boldsymbol{u}) \delta_{\alpha \beta}\bigg) -
\nonumber \\
&& - 2 \mu \bigg (\frac{\partial \Lambda_{\alpha \beta}}{\partial t} - \frac{1}{3}
\frac{\partial}{\partial t}(\boldsymbol{\nabla}\! \cdot\! \boldsymbol{u}) \delta_{\alpha \beta}\bigg) - 
\nonumber \\
&& - \tau_r \frac{\partial^2}{\partial t^2} (p_{\alpha \beta}- 
p \delta_{\alpha \beta})\bigg] \frac{\partial u_{\alpha}}{\partial r_{\beta}} 
\approx
\nonumber \\
&& \approx \frac{\partial p}{\partial t} (\boldsymbol{\nabla}\! \cdot\! \boldsymbol{u}) - \tau_r
\frac{\partial^2 p_{\alpha \beta}}{\partial t^2} \frac{\partial u_{\alpha}}
{\partial r_{\beta}} + \tau_r \frac{\partial^2 p}{\partial t^2} (\boldsymbol{\nabla}\! \cdot \!\boldsymbol{u}) = 
\nonumber \\
&& = \frac{\partial p}{\partial t} (\boldsymbol{\nabla}\! \cdot\! \boldsymbol{u}) + \tau_r 
\frac{\partial^2 p}{\partial t^2} (\boldsymbol{\nabla}\! \cdot\! \boldsymbol{u}) -
\nonumber \\
&& - \tau_r \frac{\partial}{\partial t} \bigg (\frac{\partial p}{\partial t} (\boldsymbol{\nabla}\! \cdot\! \boldsymbol{u}) - \tau_r
\frac{\partial^2 p_{\alpha \beta}}{\partial t^2} \frac{\partial u_{\alpha}}
{\partial r_{\beta}} + \tau_r \frac{\partial^2 p}{\partial t^2} (\boldsymbol{\nabla}\! \cdot\! \boldsymbol{u})\bigg) \approx
\nonumber \\
&& \approx \frac{\partial p}{\partial t} (\boldsymbol{\nabla}\! \cdot\! \boldsymbol{u}),
\label{eq:bal_13}  
\end{eqnarray}
where we use the same above mentioned rules of the first order approximation.
Then we obtain the hyperbolic hydrodynamic balance equations of the first order
\begin{eqnarray}
\frac{\partial \rho}{\partial t} + \nabla (\rho \boldsymbol{u}) =0,
\label{eq:bal_14}
\end{eqnarray}
--- the equation of continuity,
\begin{eqnarray}
&& \bigg (\tau_r \frac{\partial}{\partial t} + 1\bigg) \bigg [\rho \bigg (\frac{\partial \boldsymbol{u}}
{\partial t} + (\boldsymbol{\nabla}\!\cdot\!\boldsymbol{u}) \boldsymbol{u}\bigg)\bigg]=
\nonumber \\
&& = -\boldsymbol{\nabla} \bigg [\bigg (\tau_r \frac{\partial}{\partial t} + 1\bigg)p - \frac{\mu}{3} (\boldsymbol{\nabla}\!\cdot\!\boldsymbol{u})\bigg] + \mu \nabla^2 \boldsymbol{u}
\label{eq:bal_15}
\end{eqnarray}
--- the hyperbolic Navier-Stokes equation,
\begin{eqnarray}
&& \bigg (\tau_r \frac{\partial}{\partial t} + 1\bigg) \bigg [\rho \bigg (\frac{\partial T}
{\partial t} + (\boldsymbol{\nabla}\! \cdot\! \boldsymbol{u}) T\bigg)\bigg]=
\nonumber \\
&& = \frac{m\lambda}{c_v k} \nabla^2 T - \frac{1}{c_v} (\boldsymbol{\nabla}\! \cdot\! \boldsymbol{u})\bigg (\tau_r \frac{\partial}{\partial t} + 1\bigg) (\rho T)
\label{eq:bal_16}
\end{eqnarray}
--- the hyperbolic heat conduction equation.
Here, $c_v=3/2$ and $p=nkT$.

If $\boldsymbol{u}=0$ then Eq. (\ref{eq:bal_16}) takes the form:
\begin{eqnarray}
\tau_r \frac{\partial^2 T}{\partial t^2} + \frac{\partial T}
{\partial t} = \frac{m \lambda}{c_v k \rho} \nabla^2 T,
\label{eq:bal_17}
\end{eqnarray}
which is the hyperbolic heat conduction equation in static medium.

Equations (\ref{eq:bal_14})-(\ref{eq:bal_17}) are transformed to the classical hydrodynamic balance equations if $\tau_r=0$. 

\section{HYPERBOLIC HEAT FLUX AND HYPERBOLIC HEAT CONDUCTION EQUATION}

\label{sec:hhf}

In this section we consider the heat propagation in static medium, i.~e. when the mass velocity equal zero $\boldsymbol{u}=0$, based on the results obtained for the heat flux and the heat conduction equation in the previous section.

One can see that Eq. (\ref{eq:bal4}) is the hyperbolic heat flux Eq. (\ref{eq:grad}) where the hyperbolic relaxation time $\tau_g$ is the Maxwellian relaxation time $\tau_r$:
\begin{equation}
q_{\alpha} = - \lambda \frac{\partial T}{\partial 
r_{\alpha}} - \tau_r \frac{\partial 
q_{\alpha}}{\partial t},
\label{eq:hhf1}
\end{equation}
where the thermal conductivity $\lambda$ is defined by Eq. (\ref{bal6a}). Thus, we show that the Cattaneo equation (\ref{eq:grad}) is the result of the relaxation time approximation.

In the limit $\tau_r \to 0$ we obtain the classic form for the heat flux
\begin{equation}
q_{\alpha}^0 = \lim_{\tau_r \to 0} q_{\alpha} = - \lambda \frac{\partial T}{\partial r_{\alpha}},
\label{eq:hhf3}
\end{equation}
which corresponds to the case when $t \gg \tau_r$ (cp. Eq. (\ref{eq:fa9a})).

Let us write the hyperbolic heat conduction equation or telegraph equation (\ref{eq:bal_17}):
\begin{equation}
\tau_r \frac{\partial^2 T}{\partial t^2} + \frac{\partial T}
{\partial t} = a \nabla^2 T,
\label{eq:hhc1}
\end{equation}
where
\begin{equation}
a = \frac{m \lambda}{c_v k \rho}
\nonumber
\end{equation}
is the thermal diffusivity. Taking into account that the specific heat is $C = c_v k/m$, the thermal diffusivity is written as
\begin{equation}
a = \frac{\lambda}{C \rho},
\label{hhc1a}
\end{equation}
which is in agreement with the classic definition \cite{jou96,ferziger72,kerson63}.

Denoting $w^2=(5/6) \bar{v}^2$ and substituting $\lambda$ from (\ref{bal6a}) we obtain the expression for the $a$:
\begin{equation}
a = \tau_r w^2. 
\label{eq:hhc2}
\end{equation}

From Eq. (\ref{eq:hhc1}) one can see, that for $\tau_r \to 0$ we have the parabolic heat conduction equation 
\begin{equation}
\frac{\partial T}{\partial t}= a \frac{\partial^2
T}{\partial r_{\alpha}^2}. \label{eq:hhc4}
\end{equation} 

For $\tau_r \to \infty$ the expression (\ref{eq:hhc1}) becomes a wave equation 
\begin{equation}
\frac{\partial^2 T}{\partial t^2} = w^2 \frac{\partial^2 
T}{\partial r_{\alpha}^2}, \label{eq:hhc5}
\end{equation}
whose solutions are known in the literature as the second sound, which was observed in solids at low temperatures \cite{jou96}.

The relaxation time $\tau_r$ is the time of establishing of the local equilibrium Maxwell distribution (local thermodynamical equilibrium) by the definition from Eq. (\ref{eq:if_0}). The exact value for the $\tau_r$ depends on the type of the molecular interaction, i.e. on the collision integral (cp. Eqs. (\ref{eq:be2}) and (\ref{eq:if_0})). It is known that the relaxation time $\tau_r$ and the mean free time $\tau_0$ have the same order. However, let us estimate it on the other hand.

As follows from structure of Eq. (\ref{eq:hhc1}) the heat front is propagated  with the speed $v_f$ \cite{jou96,sobolev91,lykov67,skryl00a}:
\begin{equation}
v_f = \sqrt{\frac{a}{\tau_r}} = \sqrt{\frac{5}{6}} \bar{v} \approx \bar{v},
\label{eq:hhc6}
\end{equation}
i.e. practically with the average molecular velocity.
If we suppose that the distance of the propagation of the heat front for the time $\tau_r$ is not less than the mean free path
\begin{equation}
\tau_r v_f \geqslant \tau_0 \bar{v},
\label{eq:hhc7}
\end{equation}
the estimation for the relaxation time has the form
\begin{equation}
\tau_r \geqslant \sqrt{\frac{6}{5}} \tau_0.
\label{eq:hhc8}
\end{equation}
One can see that the expression (\ref{eq:hhc8}) is not in distinction with the known fact that the  equilibrium Maxwell distribution is established over 3-4 molecular collisions \cite{ferziger72,kerson63}. In this way, the equations (\ref{eq:bal_15}) - (\ref{eq:bal_17}) are suitable for the description of the processes at time intervals $t$ which have order of the mean free time $\tau_0$ for molecules. The limit $\tau_0 \to 0$ corresponds to a process when $t \gg \tau_0 $, i.~e. when the processes described by the classical Navier-Stokes and heat conduction equations were already formed.

In conclusion of this section we estimate the relaxation times for Nitrogen and phonons in Aliminium. The expression for the Maxwellian relaxation time of mono-atomic gas is very simple (see Eq. \ref{bal5a}):
\begin{equation}
\tau_r = \frac{\mu}{p},
\label{eq:est1}
\end{equation}
where $p$ is the static pressure. The relaxation time for phonons one can obtain in the approximation of phonon gas \cite{kittel56}:
\begin{equation}
\lambda=\frac{1}{3} C \rho u_s^2 \tau_r,
\label{eq:est2}
\end{equation}
where $u_s$ is the sound speed. Substituting $\mu$=4.15~Poise for Nitrogen at 1000 K and atmospheric pressure \cite{hanley73}, and $\lambda$=209.3~W/m K, $C$=880~J/kg, $\rho$=2.7~g/cm$^3$ and $u_s$=5.8~km/s for Aluminium \cite{koshkin76} we obtain for $\tau_r$ the values 4.096$\times$10$^{-10}$~s and 1.024$\times$10$^{-11}$~s for Nitrogen and Aluminium, respectively. As one can see from the comments to Eq. (\ref{eq:grad}) these values for the Maxwellian relaxation times are in a good agreement with the relaxation times for the hyperbolic heat conduction given by Lykov \cite{lykov67}.

\section{CONCLUSION}

\label{sec:con}

In the present work, we have applied the integral form for the Boltzmann equation and the relaxation time approximation for the collision integral. In the first order iteration we have derived the hyperbolic-like forms for the heat transfer and for the Navier-Stokes equations. For the static medium we have obtained the Cattaneo heat flux and the well known telegraph equation for the heat conduction. It was shown that the relaxation time for the hyperbolic equations is the Maxwellian relaxation time. The obtained equations can be used for the description of the fast heat-mass transfer processes of the order of the mean free time.

Additionally, for the heat propagation in static medium, we estimate  the hyperbolic and Maxwellian relaxation times for the heat conduction in Nitrogen and Aluminium. The obtained values are found to be in a good agreement with experiment.

\appendix


\section{Calculation of $M(t,t)$ and $L(t,t)$}
\label{sec_app2}

\subsection{}

Let us write the expression (\ref{eq:fa10}) for $M(t)=M(t,t)$:
\begin{equation}
M(t)=\frac{1}{10} \int m c_\alpha c_\beta \bigg [\frac{f^{(0)}}{RT}
\bigg (c_\alpha c_\beta - \frac{1}{3} c^2 \delta_{\alpha \beta}\bigg) \bigg] d^3 c,
\label{eq:app2a_1}
\end{equation}
\begin{equation}
M(t)=\frac{m}{10 R T} \int f^{(0)} \bigg [
(c_\alpha c_\beta)(c_\alpha c_\beta) - \frac{1}{3} c^2 (c_\alpha c_\beta)
\delta_{\alpha \beta}) \bigg] d^3 c.
\label{eq:app2a_2}
\end{equation}

For the expression in brackets we have: 
\begin{equation}
(c_\alpha c_\beta)(c_\alpha c_\beta) - \frac{1}{3} c^2 (c_\alpha c_\beta)
\delta_{\alpha \beta}) = \frac{2}{3} \sum_{\beta=1}^3 \sum_{\alpha=1}^3 
c_{\alpha}^2 c_{\beta}^2.
\label{eq:app2a_3}
\end{equation}

Then $M$ takes the form:
\begin{equation}
M(t)=\frac{m}{10 R T} \frac{2}{3} \sum_{\beta=1}^3 \sum_{\alpha=1}^3 
\int f^{(0)} c_{\alpha}^2 c_{\beta}^2 d^3 c,
\label{eq:app2a_4}
\end{equation}
and it can be separated into two parts:
\begin{equation}
M(t)=3 J_{\alpha \alpha} + 6 J_{\alpha \beta},
\label{eq:app2a_5}
\end{equation} 
where $J_{\alpha \alpha}$ and $J_{\alpha \beta}$ have simple expressions:
\begin{equation}
J_{\alpha \alpha}=\frac{m}{10 R T} \frac{2}{3} \int f^{(0)} 
c_{\alpha}^4 d^3 c = \frac{1}{5} n k T,
\label{eq:app2a_6}
\end{equation}
\begin{equation}
J_{\alpha \beta} = \frac{m}{10 R T} \frac{2}{3} \int f^{(0)} 
c_{\alpha}^2 c_{\beta}^2 d^3 c = \frac{1}{15} n k T.
\label{eq:app2a_7}
\end{equation}

Finally, $M$ has the form:
\begin{equation}
M(t) = n k T
\label{eq:app2a_8}   
\end{equation}

\subsection{}

Let us write now the expression (\ref{eq:fa11}) for $L(t)=L(t,t)$:
\begin{equation}
L(t)=\frac{1}{3} \int c_{\alpha} \frac{m c^2}{2} 
\frac{f^{(0)}}{T} 
c_{\alpha} \bigg (\frac{m}{2 kT} c^2 - \frac{5}{2}\bigg) d^3c.
\label{eq:b1}
\end{equation}
Then we integrate on all space of velocities and obtain
\begin{eqnarray} 
&&L(t)= \frac{m}{6 T} n (\frac{m}{2 \pi k T})^{\frac{3}{2}} 
\int \int \int c^4 e^{-A c^2} \times \nonumber \\
&& \times \bigg (\frac{m}{2 kT} c^2 - \frac{5}{2}\bigg) dc_x dc_y dc_z 
= \frac{5}{2} \frac{k T}{m} n k. \label{eq:b2}
\end{eqnarray}

Taking into account the expression for the average molecular velocity $\bar{v}^2 = 2 k T/m$, one can write: 
\begin{equation}
L(t) = \frac{5}{4} \bar{v}^2 n k. \label{eq:b3}
\end{equation}

\end{document}